\def\cps{s$^{-1}$}

\documentstyle[aaspp4]{article}

\begin{document}

\title{An X-Ray Dip in the X-Ray Transient 4U~1630-47}

\authoremail{jtomsick@phys.columbia.edu}

\author{J.A. Tomsick, I. Lapshov, P. Kaaret}
\affil{Department of Physics and Columbia Astrophysics Laboratory, \
Columbia University, 538 W. 120th Street, New York, NY 10027}

\begin{abstract}

An x-ray dip was observed during a 1996 \it Rossi X-Ray Timing Explorer \rm
observation of the recurrent x-ray transient 4U~1630-47.  During the dip, 
the 2-60~keV x-ray flux drops by a factor of about three, and, 
at the lowest point of the dip, the x-ray spectrum is considerably 
softer than at non-dip times.  We find that the 4U~1630-47 dip is best 
explained by absorption of the inner part of an accretion disk, while the 
outer part of the disk is unaffected.  The spectral evolution 
during the dip is adequately described by the variation of a single parameter, 
the column density obscuring the inner disk.

\end{abstract}


\section{Introduction}

Dips have been observed in the x-ray lightcurves of about twenty x-ray 
binaries.  The characteristics of the dips vary from source to source 
requiring a number of different explanations for the dips.  In many cases, 
dips are explained by absorption of the x-ray source by material passing 
in front of the source (White, Nagase, and Parmar 1995).  In some sources, 
dips may be due to accretion disk instabilities, which cause the ejection 
of the inner part of the disk 
(Belloni et al. 1997; Greiner, Morgan, and Remillard 1996).

Here, we present an analysis of an x-ray dip observed in the 1996 outburst of 
4U~1630-47.  The transient x-ray source 4U~1630-47 has been identified as a black hole
candidate based on the shape of its x-ray spectrum and the lack of type I x-ray
bursts (Parmar, Stella, and White 1986).  4U~1630-47 is unusual among x-ray 
transients due to the fact that its outburst recurrence time is relatively 
short and regular staying between 600 and 700 days (Kuulkers et al. 1997b).  Although 
optical observations of 4U~1630-47 have been made (Parmar et al. 1986), an optical 
companion has not been identified.  It has been suggested that 4U~1630-47 may 
be a superluminal radio jet source similar to GRS~1915+105 and GRO~J1655-40 based 
on its timing properties (Kuulkers, van der Klis, and Parmar 1997a).  For this reason, 
it is interesting to compare the properties of the 4U~1630-47 dip to the dips in 
GRS~1915+105.

In this paper, \S 2 includes a description of the 4U~1630-47 
observations and data analysis.  In \S 3, the results of the analysis are presented, 
and two possible causes for the dip are considered.  We find that the dip can be 
explained by absorption of part of the x-ray source.  In \S 4, we discuss the 
implications of the dip.

\section{Observations and Analysis}

The x-ray transient 4U~1630-47 was observed with the \it Rossi X-Ray Timing
Explorer \rm (RXTE) (Bradt, Rothschild, and Swank 1993) several times 
during an outburst which began in March 1996 and lasted for 115~days 
(Levine et al. 1996a).  In addition to the continual coverage provided by the 
All-Sky Monitor (ASM) (Levine et al. 1996b), 15 pointed observations were made 
between UTC~May~3,~1996~20:49:34 and UTC~June~4,~1996~20:22:13 with RXTE.  The 
typical on-source time for a pointed observation is 3000~s.  Proportional Counter 
Array (PCA)(Jahoda et al. 1996) 2-60~keV lightcurves were produced 
with 0.5~s time bins for all of the pointed observations of 4U~1630-47 in order to 
look for x-ray dips.  Based on a visual inspection of these lightcurves, we conclude 
that the only x-ray dips which occurred during the pointed observations were those 
observed during the May~3 observation.  In this paper, we focus on data from the PCA 
for the first 1200~s of the pointed observation beginning UTC May 3, 1996 20:49:34.

The PCA lightcurves for the 1200~s following UTC May 3, 1996 20:49:34 
(MJD 50206.8679188), shown in Figure~1, were produced using
B\_2ms\_16A\_0\_35\_Q mode data which provides 16 energy bins from 2-13.1~keV 
and 0.001953125~s time resolution.  The lightcurves include the combined count
rates for five PCUs and are background subtracted.  During the May~3 
observation, the non-dip source flux is about 240~mCrab (2-60~keV) 
corresponding to a flux of $7.0\times10^{-9}\rm~erg~cm^{-2}~s^{-1}$. 

Energy spectra have been produced using Standard 2 PCA data and 
the version 2.02 response matrix.  This response matrix gives much 
better fits to the Crab for PCUs 0, 2, 3, and 4 than for PCU 1.  Also, based on 
spectral fits to the Crab, the response matrix is much better for the top xenon 
layer than for the other two xenon layers.  Thus, only data from the top xenon 
layers of PCUs 0, 2, 3, and 4 has been used in making energy spectra.  A systematic 
error of 1\% is assumed to account for uncertainties in the response matrix 
(Jahoda et al. 1997).  Background subtraction has been performed 
including estimates for both the particle and the x-ray background (Stark et al. 1997).  
Spectral fits have been calculated using the XSPEC software (Shafer et al. 1991).

\section{Results}

Figures 1a, 1b, and 1c show lightcurves binned in 1~s intervals for three 
energy bands:  2-3.8~keV (soft band), 3.8-5.7~keV (middle band), and 
5.7-13.1~keV (hard band).  We identify four dipping events in the x-ray 
lightcurve, which are labelled in Figure 1a:  a short dip with a duration 
of $\sim$6~s labelled ``1", a longer dip lasting $\sim$140~s labelled ``2", 
and two more short dips labelled ``3" and ``4".  During dip 2, the 2-60~keV 
x-ray flux decreases by a factor of three.  Here, we consider two possible
causes for the dips:  absorption or ejection of the inner part of the disk
via an accretion disk instability.

Exponential fits to the lightcurves during ingress and egress of dip 2 have 
been calculated in order to compare the e-folding times for the three energy 
bands.  For the ingress fits, 23 seconds of data after the start of dip 2 
was used, and for the egress fits, 57 seconds of data preceeding the
end of dip 2 was used.  Table 1 shows the e-folding times and the values of 
$\chi_\nu^{2}$ for the six fits.  For both ingress and egress, 
there is a significant increase in e-folding time with increasing energy,
suggesting that dip 2 is due to absorption by an object which has a lower 
column density at its edges than in its interior.  This result does not 
appear to be consistent with dip 2 being caused by ejection of the inner 
disk since it is difficult to see how the soft x-ray emitting region could 
be ejected before the hard x-ray emitting region.  Also, it is not clear
why the hardest part of the x-ray source would recover first at the end 
of the dip.  We also find that the egress e-folding times are substantially 
longer than the ingress e-folding times.  If absorption is the cause of the
dip, this result may indicate an asymmetry in the density or the shape of 
the absorber.

Figure 2 shows the relationship between spectral hardness and count rate.  
The hardness ratio is constant above about 1500~\cps.  
However, as the count rate decreases from its non-dip level, spectral changes 
become apparent.  For intermediate count rates (1000~\cps~to~1500~\cps), the 
spectrum is significantly harder than in its non-dip state, and for 
count rates below 1000~\cps, the spectrum softens.  The hardness increase at 
intermediate count rates is consistent with absorption, but inconsistent with
ejection of the inner part of the accretion disk.  The softening of the spectrum 
at count rates below 1000~\cps~can be explained by ejection of the inner part 
of the disk, but cannot be explained by a cold absorber covering the entire 
x-ray source.  However, the softening is consistent with the presence of an 
extended soft x-ray source and provides motivation for a two component model 
where one component is absorbed during the dip and a second, softer, extended 
component is not absorbed during the dip.  

For spectral analysis, dip and non-dip data have been selected 
from the May~3 observation based on the source intensity during each of the 
16~s integrations provided by Standard 2 data.  The dip spectrum 
consists of a 64~s integration and 960~s of data are used for the non-dip 
spectrum.  For the non-dip spectrum, single component models do not provide 
acceptable fits:  a powerlaw fit gives $\chi_\nu^{2}=14$ for 41 
degrees of freedom (dof), a fit using a 
thermal bremsstrahlung model gives $\chi_\nu^{2}=14$ for 41 dof, a blackbody 
disk model gives $\chi_\nu^{2}=38$ for 41 dof, and a cut-off powerlaw gives 
$\chi_\nu^{2}=4.3$ for 40 dof.  An acceptable fit to the energy spectrum is 
achieved using a model combining a disk-blackbody component 
(Makishima et al. 1986) with a powerlaw component 
($\chi_\nu^{2}=0.68$ for 39 dof).  This is consistent with fits to 
spectra from previous observations of 4U~1630-47 (Parmar et al. 1997).

Since the data suggests the presence of a soft component that is not absorbed
during the dip, we consider a model where flux coming from the outer region of 
the disk is not affected during the dip and the inner region of the disk is 
absorbed during the dip.  The radius dividing the unabsorbed and absorbed 
regions of the disk is one of the free parameters in the model.  
The sharp cutoff between the unabsorbed and absorbed regions is an
approximation of the actual physical system.  However, the data are not of
sufficient quality to distinguish between this model and a more complex
model.  Since it is not known 
where the powerlaw component originates, the model allows for part of the 
powerlaw component to come from the unabsorbed region and part to 
come from the absorbed region, but it is assumed that the powerlaw index 
($\alpha$) is the same in both regions.

The disk model used here is the disk-blackbody model discussed in 
Makishima et al. (1986).  Using this model, we are able to break the disk into
two pieces and absorb the spectra from the two pieces differently.  For each
piece, the unabsorbed spectrum can be written as,
\begin{equation}
f_{disk}(x_{1},x_{2},N,T_{max};E) = C_{0}N\int_{x_{1}}^{x_{2}} xB[T(x);E]dx
\end{equation}
where $B[T(x);E]$ is the blackbody flux per unit photon energy from a unit
surface area of temperature $T(x)$,  $x$ is the radial distance from the
compact object in units of the inner radius of the disk ($r_{in}$), and $x_{1}$
and $x_{2}$ are, respectively, the inner and outer radii for the piece of the disk 
in units of the inner radius of the disk.  The expression used for T(x) is given in
equation 3.23 of Pringle (1981).  $C_{0}$ is a constant equal to 
$2\pi(1\rm~km/10\rm~kpc)^{2}$.  Given $x_{1}$ and $x_{2}$, the 
spectrum is determined by the two parameters $N$ and $T_{max}$.  $N$, the 
normalization parameter, is given by 
$((r_{in}/1\rm~km)/(d/10\rm~kpc))^{2}~$cos~$i$, where $d$ is the 
distance to the source and $i$ is the disk inclination.  $T_{max}$ is equal to 
the disk temperature at a radius $(49/36)~r_{in}$ (Pringle 1981).  
Combining the disk component with the powerlaw component and including 
absorption gives, 
\begin{equation}
f(x_{1},x_{2},N,T_{max},A,\alpha,N_{\rm H};E) = \
e^{-\sigma(E)N_{\rm H}}[f_{disk}(x_{1},x_{2},N,T_{max};E)+ \
A \left(\frac{E}{1~\rm keV \it}\right)^{-\alpha}]
\end{equation}
for the spectrum from the region between $x_{1}$ and $x_{2}$.  The 
cross section ($\sigma(E)$) includes both photoelectric absorption 
(Morrison and McCammon 1983) and Compton scattering.  Combining 
the emission from the absorbed and unabsorbed regions gives,
\begin{equation}
S = e^{-\sigma(E)N_{\rm H}}f(1.0,\frac{r_{d}}{r_{in}},N,T_{max},A_{in},\alpha,N_{\rm H}^{(is)};E)+f(\frac{r_{d}}{r_{in}},\frac{r_{out}}{r_{in}},N,T_{max},A_{out},\alpha,N_{\rm H}^{(is)};E)
\end{equation}
where $N_{\rm H}^{(is)}$ is the column density for interstellar absorption, 
$N_{\rm H}$ is the column density for the extra absorbing material obscuring
the inner disk, and $r_{d}$ is the radius dividing the absorbed and unabsorbed regions.
We assume that $N_{\rm H}$ is the same for photoelectric absorption and
Compton scattering.  The parameter $r_{out}/r_{in}$ is fixed so that the 
x-ray flux from disk radii beyond $r_{out}$ is negligible.  In all, there are eight 
free parameters in the model.  The unabsorbed emission from the disk is determined 
by $N$ and $T_{max}$.  The parameters $A_{in}$, $A_{out}$, and $\alpha$ specify 
the emission due to the powerlaw component.  The other free parameters are 
$N_{\rm H}^{(is)}$, $N_{\rm H}$, and $r_{d}/r_{in}$.

This model has been used to simultaneously fit the dip and non-dip spectra.  
In calculating the fit, the only parameter which is allowed to be 
different for the two spectra is the column density ($N_{\rm H}$) of the absorbing 
material.  This model provides a good fit to the two spectra with 
$\chi_\nu^{2}=0.65$ for 79~dof.  The fit parameters with 
their 68\% confidence errors are given in Table~2 and the fitted spectra are
shown in Figure~3.  The extra material absorbing the inner region of the disk is 
found to have a column density of 
$N_{\rm H} = (86.8\pm 4.4)\times10^{22}\rm~cm^{-2}$ during the dip and 
$N_{\rm H} = (7.32\pm 1.47)\times10^{22}\rm~cm^{-2}$ during the non-dip, 
and the interstellar absorption is 
$N_{\rm H}^{(is)} = (8.26\pm 0.74)\times10^{22}\rm~cm^{-2}$.  
The non-dip and dip spectra cannot be simultaneously fit with a model where 
$N_{\rm H}$ is set to zero for the non-dip and $N_{\rm H}^{(is)}$ is the same
for the inner and outer regions of the disk.

Past observations of 4U~1630-47 show that it is reasonable to think that
some portion of the absorption measured in the non-dip spectrum is due to 
material close to the source since the column density has shown a high degree 
of variability from observation to observation.  Column densities of
$(6.4\pm 0.2)\times10^{22}\rm~cm^{-2}$, 
$(9.51\pm 0.11)\times10^{22}\rm~cm^{-2}$, and 
$(14.3\pm 1.2)\times10^{22}\rm~cm^{-2}$ were measured in three separate 
observations (Parmar, Angellini, and White 1995, Parmar et al. 1997).  
It would be reasonable to assume that the lowest of these measurements provides 
an upper limit on the interstellar absorption, which would make our value of
$N_{\rm H}^{(is)}$ only slightly higher than expected.

Above about 13~keV, the powerlaw component dominates the non-dip and dip 
spectra and the disk-blackbody component is negligible.  Figure~3 shows
that the dip flux is lower than the non-dip flux at all energies including
the powerlaw dominated region.  Thus, if the dip is caused by
absorption, then at least part of the powerlaw component must be absorbed.
Comparing the values of $A_{in}~(78.1\pm~30.4)$ and $A_{out}~(2.7\pm~4.7)$ 
indicates that most of the powerlaw component comes from the absorbed region.  
Further evidence that most of the powerlaw component
is absorbed is the fact that if $A_{out}$ is fixed to zero, and the 
simultaneous fit is recalculated, the quality of the fit does not change 
($\chi_\nu^{2}=0.65$ for 80~dof).  

If the only parameter that changes with time is the column density of the absorber, 
a measured count rate in a particular energy band corresponds to a unique value 
of $N_{\rm H}$, and it is possible to see if variations in the column density
can produce the features observed in the lightcurves shown in Figure~1.  
Figure~4 shows the column density estimates derived from the count rates
in the middle and hard bands during the dip.  There is good agreement between 
the $N_{\rm H}$ values calculated from the two bands.  The model predicts almost 
the same flux in the soft 
band for all values of $N_{\rm H}$ above about $40\times10^{22}\rm~cm^{-2}$;  
thus, for column densities above this level, the soft band flux provides no 
information and is not shown in Figure~4.  The absorption model predicts that the 
drop in the column density observed in the middle of dip 2 should produce a 20\% 
increase in the soft band flux.  This is consistent with the observed 
$21.8\pm 2.7$\% flux increase.
The model assumes that the intrinsic (unabsorbed) intensity of 4U~1630-47 remains 
constant during the dips, and that the observed change in flux is due entirely to
variations in the column density.  Using the non-dip 
lightcurve with 0.5~s time bins, we find that the source variability is below 
the 5\% level over a time interval comparable to the duration of the dip (140~s)
indicating that this assumption is reasonable.

A model where an accretion disk instability leads to ejection of the inner
part of the accretion disk has been used to explain the x-ray dips observed
in GRS~1915+105.  Since this is a model which can produce spectrally soft dips, 
we have considered it as a possible mechanism for the 4U~1630-47 dip.  
Based on fits to GRS~1915+105 spectra, Belloni et al. (1997) 
found that the inner radius of the accretion disk is larger during dip times 
than during non-dip times.  We fit the 4U~1630-47 non-dip and dip spectra 
simultaneously using a disk-blackbody plus powerlaw model and requiring the absorption
(column density and absorption edge parameters) to be the same for the non-dip and 
dip spectra.  For the dip spectrum, 
the disk-blackbody normalization is consistent with zero ($26\pm 36$), and
the powerlaw is the only significant spectral component.  This could be 
interpreted as meaning that the entire x-ray emitting portion of the disk 
is ejected during the dip.  However, large residuals are observed near 
7~keV for the dip spectrum but not for the non-dip spectrum.  This suggests 
the presence of an iron K-absorption edge only in the dip spectrum, which 
supports absorption as the cause of the dip.

\section{Discussion}

The spectrally soft x-ray dips observed in 4U~1630-47 are best explained 
by a model where part of the x-ray source is absorbed, while another part 
of the x-ray source is unaffected during the dips.  The models used to explain 
absorption dips in other sources typically place the absorbing material in the 
accretion disk or in a wind from a high mass companion.  In the former 
case, the dips are probably produced by an accumulation of material where the 
stream of material from the companion impacts the disk.  This results in dips 
being observed at certain orbital phases, and most of these sources are described 
as periodic dippers.  In order for dips to be observed, the inclination angle 
must be relatively high ($i>70$~degrees) (White et al. 1995).  Strong stellar 
winds often exist in binaries where the optical companion is an O or B type 
star.  In 4U~1630-47, neither the inclination angle nor the spectral type of 
the companion are known so there is no reason to assume either of these 
possibilities a priori.  In this section, we discuss the properties of the 
dip in relation to these two possibilities.

Frank, King, and Lasota (1987) (henceforth FKL) explain the dips using a model 
where the stream of material from the companion is thicker than the scale height of
the accretion disk so that a fraction of the stream flows above and below the disk.
In this model, when the matter from the stream is irradiated by x-rays coming from 
close to the compact object, ionization instabilities cause the material to separate 
into a two-phase medium consisting of cold, relatively dense clouds, which are 
responsible for the dips, in a hot intercloud medium.  Here, we compare the cloud 
sizes and densities derived by FKL to estimates for those quantities for the absorber 
producing the dips in 4U~1630-47.  To estimate the absorber velocity, we assume that 
the absorber is in a circular orbit about a $3M_{\sun}$ compact object, where the mass 
estimate comes from assuming that the highest observed flux for 4U~1630-47 
(Parmar et al. 1995) is the Eddington luminosity at 10~kpc.  We use 
$r=4.9\times10^{10}~P_{1day}^{2/3}$~cm for the distance from the compact object to 
the absorber (FKL), where $P_{1day}$ is the binary orbital period in units of days.  
If we assume that a typical cloud caused dip~1, then the cloud diameter is 
$5.4\times10^{8}~P_{1day}^{-1/3}$~cm.  By using the dip~1 duration, we are 
assuming that dips 2 and 4 last longer and have higher peak 
column densities because they are caused by multiple clouds.  Using standard 
values for the accretion rate, the fraction of matter in the incoming stream which 
reaches $r$, and the scale height of the disk given in FKL, the cloud sizes 
expected from the model of FKL are $6.4\times10^{8}~P_{1day}^{11/9}$~cm for 
$M=3M_{\sun}$, and the cloud size estimates agree if the orbital period is 
near one day.

We can estimate the density of the absorber that produced dip~1 if we assume 
that the linear size of the cloud along the line of sight is the same as the 
linear size in the direction perpendicular to the line of sight.  For matter 
with cosmic abundances, the cloud density is found to be 
$1.6\times10^{-9}~P_{1day}^{1/3}$~g~cm$^{-3}$. Making the same assumptions 
as above, the density derived by FKL is $2.7\times10^{-9}~P_{1day}^{-7/3}$~g~cm$^{-3}$, 
and the density estimates agree if the orbital period of the system is near one day, 
which is consistent with the orbital period derived based on the sizes 
of the clumps.  We conclude that the characteristics of the dip are consistent with 
being produced by material in the accretion disk and that the model of FKL may imply 
that the orbital period of 4U~1630-47 is about a day.

Finding that 4U~1630-47 dips periodically would be evidence that the dips
are caused by material in the accretion disk.  The other systems which exhibit 
periodic dips typically have duty cycles between 10 and 30\%.  Dips have never
been observed before in 4U~1630-47, and the longest continuous observation 
was a 7 hour $EXOSAT$ observation (Parmar et al. 1986).  
Thus, assuming that x-ray dipping is not a transient phenomenon in 4U~1630-47, 
the upper limit on the duty cycle is 1\%.  
Recent observations of dips from the superluminal jet 
source GRO~J1655-40 indicate that the dips in this source also have a very small 
duty cycle, 0.2\% for 400~s dips every 2.6~days (Kuulkers et al. 1997c).  
Most of the systems which exhibit periodic dipping are neutron star x-ray 
binaries, while GRO~J1655-40 is securely identified as a black hole system 
(Orosz and Bailyn 1997).  The fact that the dip duty cycle in GRO~J1655-40 is so 
different from the neutron star systems suggests the possibility that the difference 
is related to the nature of the compact object.  The 4U~1630-47 duty cycle may be 
small because the source contains a black hole.

To fit the dip spectrum of 4U~1630-47 by absorbing the non-dip spectrum, 
it is necessary to absorb the powerlaw component.  This is in contrast to
the x-ray burster 4U~1624-49 (Church and Baluci\'{n}ska-Church 1995) for 
which the non-dip spectrum is fit with a blackbody plus powerlaw model, and
the dip spectrum can be fit by absorbing the blackbody component,
but leaving the powerlaw component unaffected.  These results probably 
indicate that the size of the region emitting the powerlaw is larger for 
4U~1624-49 than for 4U~1630-47 and may indicate that the size of the region 
emitting the powerlaw is larger for neutron star systems than for systems 
containing black holes.  It would be interesting to compare the spectral 
evolution of GRO~J1655-40 during dips to that of 4U~1630-47 and 4U~1624-49.

We have also considered the possibility that the 4U~1630-47 dip is produced by a wind
from a high mass companion.  For 4U~1630-47, the relatively sharp transitions 
between non-dip and dip states, as observed in the lightcurve, and the corresponding 
variations in the column density indicate that the material producing the dips must
be non-uniform in density and/or shape (i.e. clumpy);  however, there is evidence 
for non-uniformity in stellar winds.  Cyg~X-1 is a HMXB where dips with sharp 
ingresses and egresses ($\sim2$~s) are observed and attributed to the wind from 
the high mass companion (Kitamoto et al. 1984).  In addition, Nagase et al. (1986) 
suggests that stellar winds are likely to be turbulent and that clumping may be 
caused or enhanced by the x-ray source.  The typical column densities observed 
for dips in HMXBs are between $10\times10^{22}\rm~cm^{-2}$ and 
$50\times10^{22}\rm~cm^{-2}$
(Haberl and White 1990; Kitamoto, Miyamoto, and Yamamoto 1989), which are 
only slightly less than the maximum column density measured for 4U~1630-47.
Based on the properties of the dips observed in HMXBs, we cannot eliminate the
possibility that the 4U~1630-47 dip was caused by material in the wind of a 
high mass companion.

Distinguishing between the possible causes of the dip in 4U~1630-47 may lead
to understanding its outburst recurrence time ($t_{r}$).  It has been suggested 
that $t_{r}$ may be the orbital period of the system and that outbursts occur 
near periastron (Parmar et al. 1995).  Finding that material in a stellar 
wind was responsible for the dip would indicate the presence of a high mass 
companion and make it likely that $t_{r}$ is the orbital period.  Alternatively, 
confirmation of an orbital period near one day suggests similarities between 
4U~1630-47 and Her~X-1, which has a 1.7~day orbital period and a 35~day on/off 
cycle.  In this picture, $t_{r}$ is analogous to the 35~day period in Her~X-1, 
which is explained by accretion disk precession (Crosa and Boynton 1980).  
Kuulkers et al. (1997b) find evidence for a change in $t_{r}$ from 600 to 
690~days for the last three 4U~1630-47 outbursts, which suggests that $t_{r}$ 
is not the orbital period.  However, the behavior of the x-ray transient 
GS~0834-430 indicates that the possibility that $t_{r}$ is related to the 
orbital period should not be completely ruled out (Wilson et al. 1997).

\acknowledgements

JAT would like to thank Eric Ford and Karen Leighly for useful discussions about the 
PCA response.

%FIGURE CAPTIONS
%\clearpage
%\figcaption[fig1.ps]{$RXTE$ PCA lightcurves of 4U~1630-47 in the (a) 2-3.8~keV, 
%(b) 3.8-5.7~keV, and (c) 5.7-13.1~keV energy bands with 1~s time resolution.  The 
%starting time is UTC May 3, 1996 20:49:34.}
%\figcaption[fig2.ps]{The (3.8-5.7~keV)/(2-3.8~keV) hardness ratio vs. count rate 
%for the first 500 seconds of the lightcurves shown in Figure 1.}
%\figcaption[fig3.ps]{Dip and non-dip $RXTE$ PCA energy spectra.  The dip spectrum and
%dip residuals are marked with diamonds.  In Figure 1b, the solid lines mark
%the dip times and the dashed lines mark the non-dip times.}
%\figcaption[fig4.ps]{The filled circles are the column density estimates derived 
%from the 5.7-13.1~keV count rates, and the open circles are the column density 
%estimates from the 3.8-5.7~keV band.  The dashed lines mark the dip and non-dip 
%values of $N_{H}$.}

%FIGURES

\clearpage
\begin{figure*}[t] \figurenum{1} \epsscale{1.0} \plotone{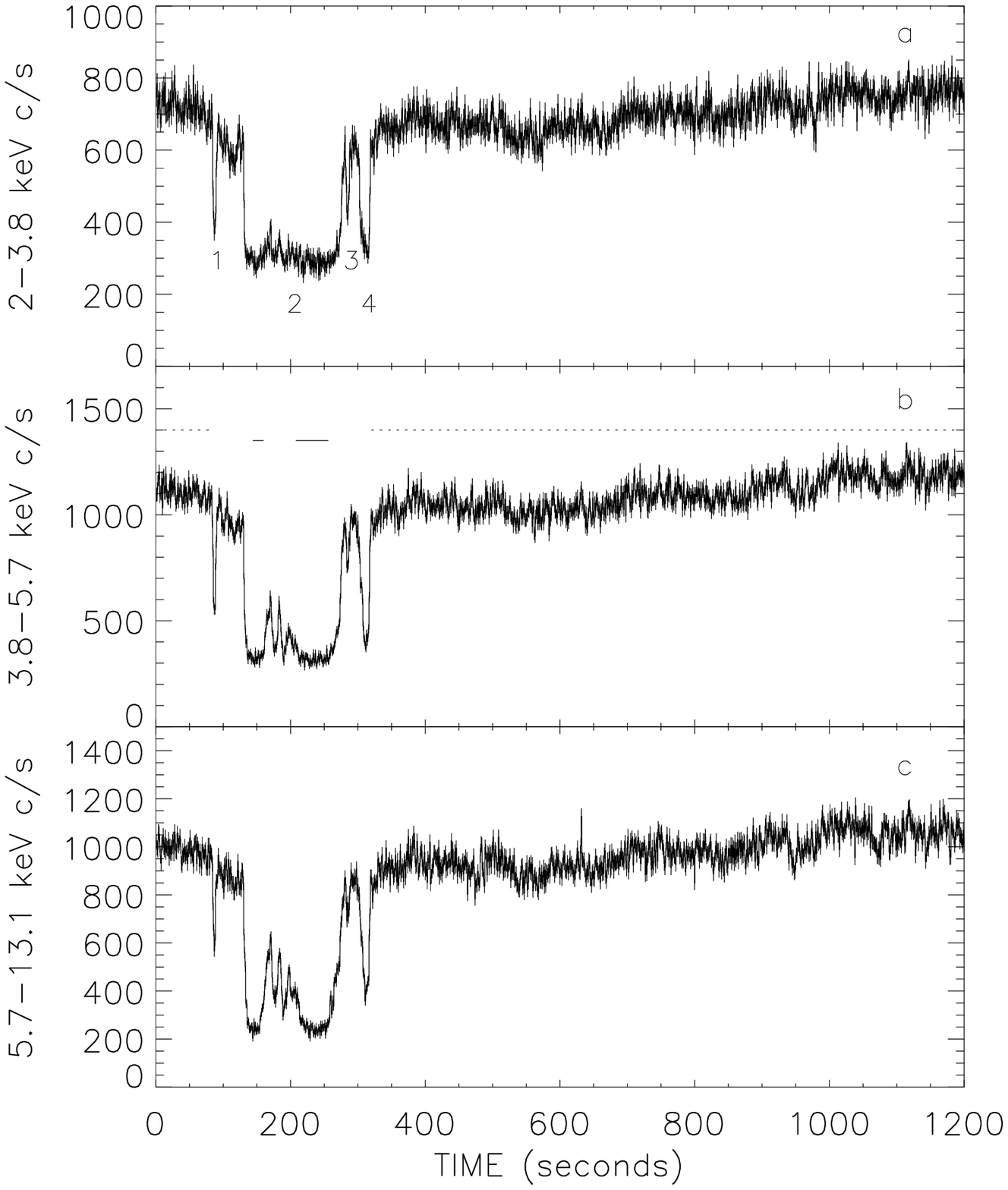}
\vspace{2.25cm}
\caption{$RXTE$ PCA lightcurves of 4U~1630-47 in the (a) 2-3.8~keV, (b) 3.8-5.7~keV, and
(c) 5.7-13.1~keV energy bands with 1~s time resolution.  The starting time is 
UTC May 3, 1996 20:49:34. \label{fig1}}
\end{figure*}

\clearpage
\begin{figure*} \figurenum{2} \epsscale{1.0} \plotone{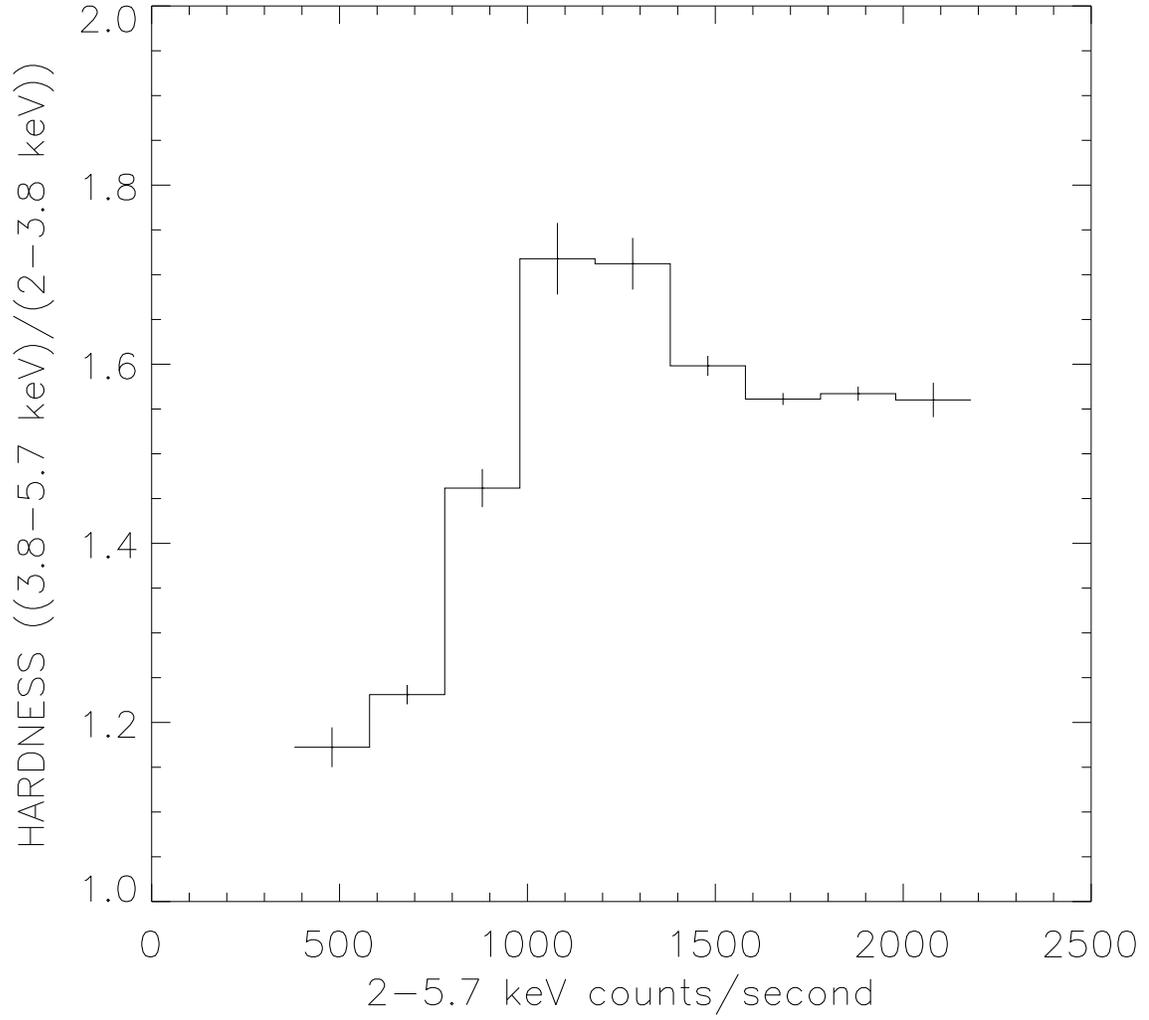}
\caption{The (3.8-5.7~keV)/(2-3.8~keV) hardness ratio vs. count rate for the first
500 seconds of the lightcurves shown in Figure 1. \label{fig2}}
\end{figure*}

\clearpage
\begin{figure*} \figurenum{3} \epsscale{0.9} \plotone{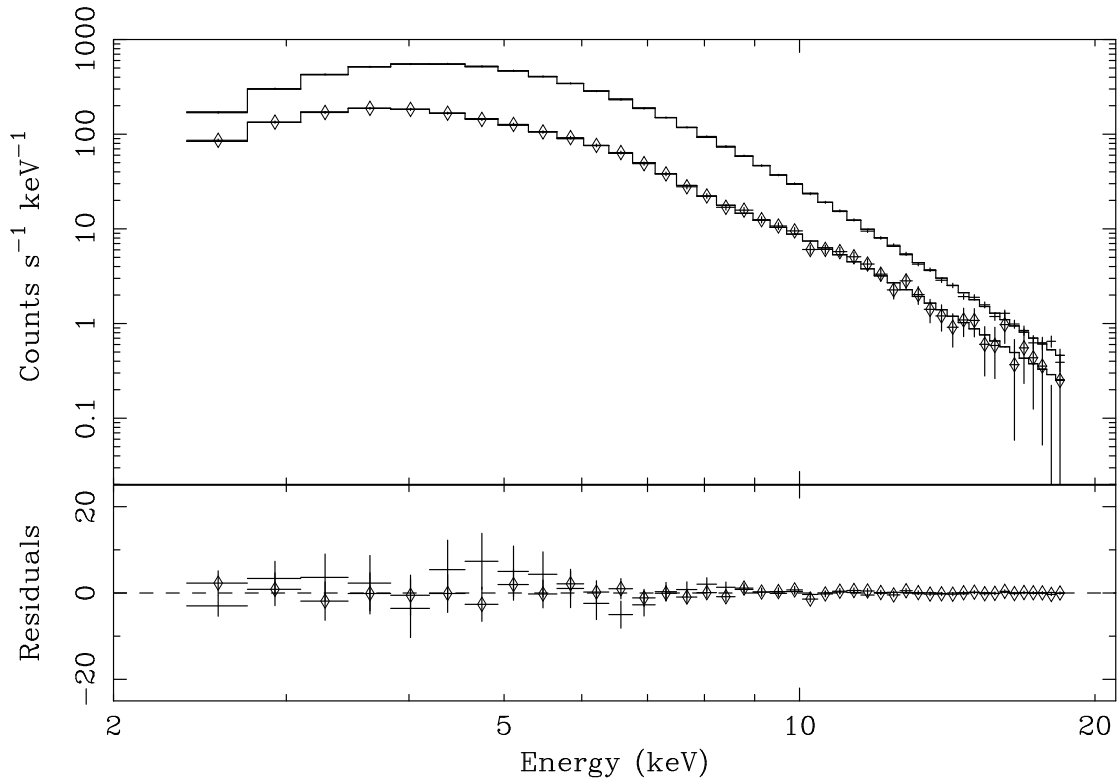}
\vspace{-3.0cm}
\caption{Dip and non-dip $RXTE$ PCA energy spectra.  The dip spectrum and
dip residuals are marked with diamonds.  In Figure 1b, the solid lines mark
the dip times and the dashed lines mark the non-dip times. \label{fig3}}
\end{figure*}

\clearpage
\begin{figure*} \figurenum{4} \epsscale{1.0} \plotone{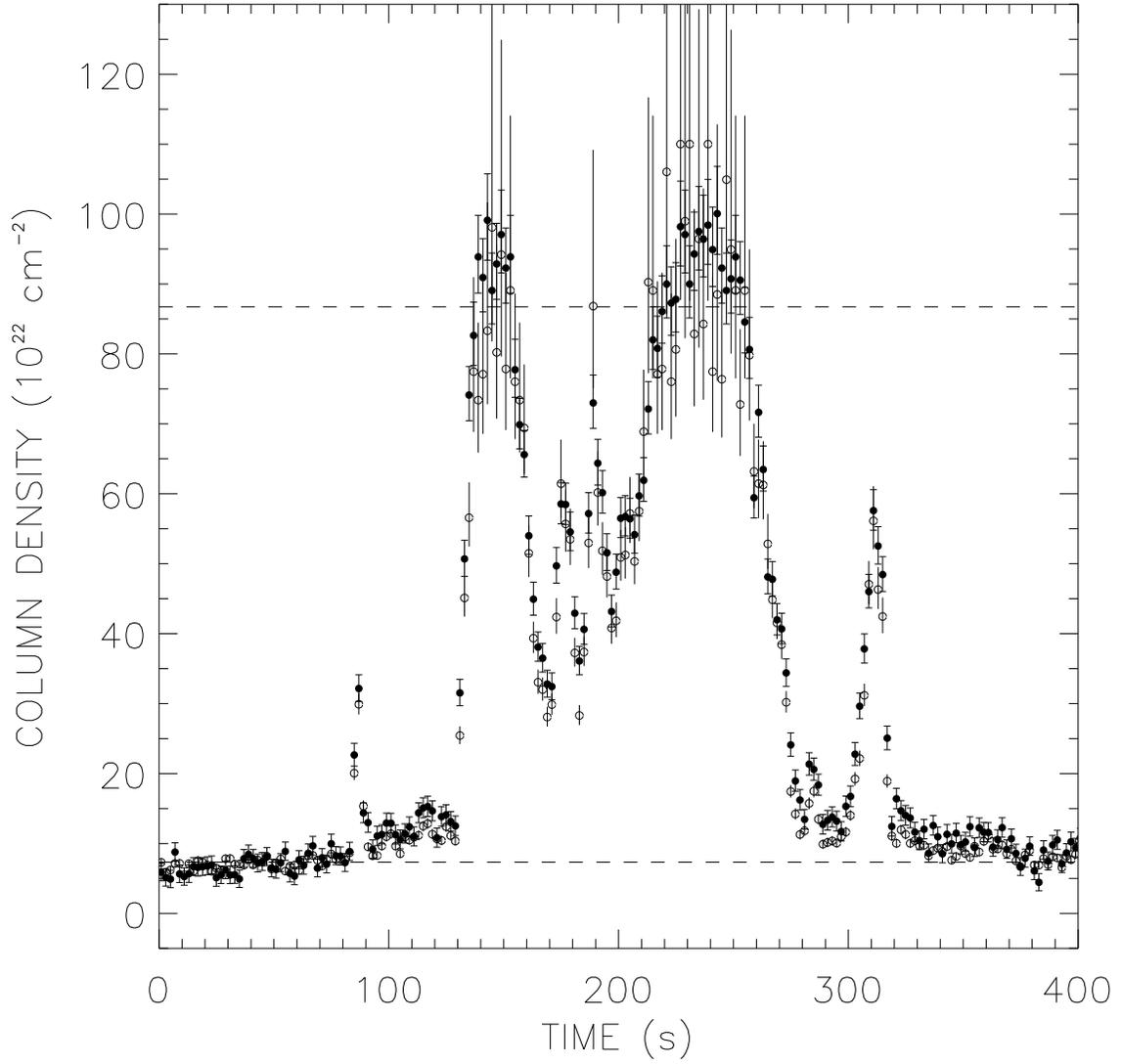}
\caption{The filled circles are the column density estimates derived from the 5.7-13.1~keV
count rates, and the open circles are the column density estimates from the 3.8-5.7~keV 
band.  The dashed lines mark the dip and non-dip values of $N_{H}$.   \label{fig4}}
\end{figure*}

%TABLES

\clearpage
 
\begin{deluxetable}{lccc}
\footnotesize
\tablecaption{Ingress and Egress e-Folding Times \label{tbl-1}}
\tablewidth{0pt}
\tablehead{
\colhead{Ingress/Egress} & \colhead{Energy Band (keV)} & \colhead{e-Folding Time (s)} & \colhead{$\chi_\nu^{2}$ (dof)}  
}
\startdata
Ingress & 2-3.8 & $1.56\pm 0.09$ & 1.013 (180)\nl
Ingress & 3.8-5.7 & $2.09\pm 0.08$ & 1.161 (180)\nl
Ingress & 5.7-13.1 & $2.78\pm 0.10$ & 1.032 (180)\nl
\hline
Egress & 2-3.8 & $4.48\pm 0.15$ & 1.158 (453)\nl
Egress & 3.8-5.7 & $6.49\pm 0.15$ & 1.075 (453)\nl
Egress & 5.7-13.1 & $10.24\pm 0.25$ & 1.199 (453)\nl
\enddata
\end{deluxetable}

\begin{deluxetable}{lcc}
\footnotesize
\tablecaption{Fit Parameters for the Simultaneous Fit to the Dip and Non-Dip Spectra \label{tbl-2}}
\tablewidth{0pt}
\tablehead{
\colhead{Parameter} & \colhead{Value} & \colhead{Dimensions}
}
\startdata
$N$ & $67.4\pm 6.7$ & dimensionless \nl
$kT_{max}$ & $1.145\pm 0.011$ & keV \nl
$A_{in}$ & $78.1\pm 30.4$ & photons cm$^{-2}$ s$^{-1}$ keV$^{-1}$ \nl
$A_{out}$ & $2.7\pm 4.7$ & photons cm$^{-2}$ s$^{-1}$ keV$^{-1}$ \nl
$\alpha$ & $4.20\pm 0.15$ & dimensionless \nl
$N_{\rm H}^{(is)}$ & $8.26\pm 0.74$ & $10^{22}\rm~H~atoms~cm^{-2}$ \nl
$r_{d}/r_{in}$ & $2.77\pm 0.10$ & dimensionless \nl
$N_{\rm H}$ (dip) & $86.75\pm 4.39$ & $10^{22}\rm~H~atoms~cm^{-2}$ \nl
$N_{\rm H}$ (non-dip) & $7.32\pm 1.47$ & $10^{22}\rm~H~atoms~cm^{-2}$ \nl
\enddata
\end{deluxetable}

\end{document}